\documentclass{ws-procs9x6}
\usepackage{wrapfig}
\newcommand{\be}{\begin{equation}}
\newcommand{\ee}{\end{equation}}
\newcommand{\ba}{\begin{eqnarray}}
\newcommand{\ea}{\end{eqnarray}}
\newcommand{\la}{\langle}
\newcommand{\ra}{\rangle}
\newcommand{\di}{ {\rm d} }

\begin{document}
\title{ \boldmath Pretzelosity distribution function $h_{1T}^\perp$}
\author{H.~Avakian,$^1\,$ A.~V.~Efremov,$^2\,$ P.~Schweitzer,$^{3,4}$ F.~Yuan$^{5,6}$}
\address{
  $^1$ 	Thomas Jefferson National Accelerator Facility,
   	Newport News, VA 23606, U.S.A. \\
  $^2$	Joint Institute for Nuclear Research, Dubna, 141980 Russia\\
  $^3$	
	Inst.\ f.\ Theoret.\ Physik II, 
        Ruhr-Universit{\"a}t Bochum, 
	D-44780 Bochum, Germany\\
  $^4$  Departement of Physics, University of Connecticut, Storrs, CT 06269, U.S.A.\\
  $^5$	RIKEN BNL Research Center, Building 510A, BNL, Upton, NY 11973, U.S.A. \\
  $^6$	Nuclear Science Division, 
	Berkeley National Laboratory, Berkeley, CA 94720, U.S.A.}

\vspace{-3mm}

\begin{abstract}
	The 'pretzelosity' distribution $h_{1T}^\perp$ is discussed. 
	Theoretical properties, model results, and perspectives to access 
	experimental information on this leading twist, transverse 
	momentum dependent parton distribution function are reviewed. 
	Its relation to helicity and transversity distributions is highlighted.
\end{abstract}
\bodymatter

\section{Introduction}
\label{Sec-1:introduction}

SIDIS allows to access information on TMDs that are defined in terms of 
light-front correlators \cite{Collins:1981uk,Collins:1984kg,Ji:2004wu,Collins:2004nx,Boer:1997nt,Mulders:1995dh}
(with process-dependent paths
\cite{Brodsky:2002cx,Collins:2002kn,Belitsky:2002sm,Cherednikov:2007tw})
\be\label{Eq:correlator}
    \phi(x,\vec{p}_T)_{ij} = \int\!\frac{\di z^-\di^2\vec{z}_T\!\!}{(2\pi)^3}\;e^{ipz}
    \la N|\bar\psi_j(0)\,{\cal W}(0,z,\mbox{path})\,\psi_i(z)|N\ra
    \Biggl|_{
	\renewcommand{\arraystretch}{0.3}
	\begin{array}{l}
	\scriptscriptstyle z^+=0\cr
	\scriptscriptstyle p^+ = xP^+ \cr
	\scriptscriptstyle \phantom{I}
	\end{array}}
    \ee
where light-cone components $p^\pm=(p^0\pm p^3)/\sqrt{2}$ are along 
the virtual photon momentum, and transverse vectors like $\vec{p}_T$ 
are perpendicular to it. 
Different TMDs are given by traces of the correlator (\ref{Eq:correlator}) 
with specific $\gamma$-matrices. There are 8 leading-twist TMDs, i.e.\
they give rise to effects that are not power suppressed in the hard scale,
and each of them contains independent information about the nucleon structure.
All leading-twist TMDs can be accessed unambiguously in SIDIS with polarized 
leptons and nucleons by observing the azimuthal distributions of produced hadrons.

In this note we will focus on the leading-twist TMD pretzelosity $h_{1T}^{\perp a}$, 
on which interesting results have been obtained from model calculations. This TMD
is responsible for a SSA $\propto\sin(3\phi-\phi_S)$ in SIDIS that could be
accessed in experiments --- most promisingly at Jefferson Lab.

\section{\boldmath Properties of $h_{1T}^{\perp a}\,$}
\label{Sec-3:What-we-know}

Let us list briefly, what we know about the pretzelosity distribution.
\begin{enumerate}
\renewcommand{\labelenumi}{\sl \Roman{enumi}.}
\item 	It can be projected out from the correlator (\ref{Eq:correlator})
	by tracing it with $i\sigma^{j+}\gamma_5$
	where it appears as the coefficient of the structure
     	$S_T^k(p_T^j p_T^k-\frac12\,\vec{p}_T^{\:2}\delta^{jk})$,
	and it has a probabilistic interpretation \cite{Mulders:1995dh}.
\item   It requires nucleon wave-function components with two units orbital 
	momentum difference \cite{Burkardt:2007rv}, and 'measures' the deviation 
	of the 'nucleon shape' from a sphere \cite{Miller:2007ae}.
        (That is why it is called 'pretzelosity'!)
\item   It is expected to be suppressed at small and large 
	\cite{Brodsky:2006hj,Burkardt:2007rv,Avakian:2007xa} $x$ 
	with respect to parton distributions like $f_1^a(x)$, $g_1^a(x)$, $h_1^a(x)$.
\item\label{item-positivity}
    	It satisfies the positivity condition \cite{Bacchetta:1999kz}
    	$2| h_{1T}^{\perp(1)a}(x)| \le f_1^a(x) - g_1^a(x)$.\\
    	The above and Soffer bound \cite{Soffer:1994ww} imply:
    	$|h_{1T}^{\perp(1)a}(x)|+ |h_1^a(x)| \le f_1^a(x)$.
\item\label{point-large-Nc}
    	In the limit of a large number of colors $N_c$ 
	in QCD it was shown that \cite{Pobylitsa:2003ty} 
	$(h_1^{\perp u}+h_1^{\perp d})/(h_1^{\perp u}-h_1^{\perp d})$
	$\sim{\cal O}(1/N_c)$ for $xN_c\sim{\cal O}(1)$ and $p_T\sim{\cal O}(1)$.
	The same pattern is predicted also for antiquarks \cite{Pobylitsa:2003ty}.
\item	It was observed in the bag model \cite{Avakian:2008dz} that 
	$h_{1T}^{\perp(1)q}(x) = g_1^q(x) - h_1^q(x)$ 
	which is confirmed in many 
        \cite{Jakob:1997wg,Avakian:2008dz,Pasquini:2008ax,pretzel-new,Meissner:2007rx}
	(but not all \cite{Meissner:2007rx,Bacchetta:2008af}) models.
\item	In simple (spectator-type) models, it has been related to chirally
    	odd generalized parton distributions \cite{Meissner:2007rx}.
\item	In SIDIS with unpolarized electrons ($U$) and transversely polarized 
	nucleons ($T$) it gives rise 
	(in combination with Collins fragmentation function 
	\cite{Collins:1992kk,Efremov:1992pe}) to an azimuthal 
	modulation of the produced hadrons proportional to $\sin(3\phi-\phi_S)$. 
	Here $\phi$ ($\phi_S$) is the azimuthal angle of the produced hadron
	(nucleon polarization vector) with respect to the virtual photon.
	The corresponding SSA is given by
	\be\label{Eq:AUT-Gauss}
    	A_{UT}^{\sin(3\phi-\phi_S)} =
    	\frac{C_{\rm Gauss}\sum_a e_a^2 x \,h_{1T}^{\perp(1) a}(x)\,
      	H_1^{\perp (1/2)a}(z)} {\sum_ae^a\,xf_1^a(x) \,D_1^a(z)}
    	\ee
	with $h_{1T}^{\perp (1)}(x)=$
	$\int\di^2\vec{p}_T\frac{\vec{p}_T^{\:2}}{2M_N^2}h_{1T}^\perp(x,\vec{p}_T^{\:2})$
	the 'transverse moment' of pretzelosity.
	Unless the DIS counts are weighted with adequate powers of
	transverse hadron momenta \cite{Boer:1997nt}
	it is necessary to assume a model for the distribution of transverse
	parton momenta. In (\ref{Eq:AUT-Gauss}) the Gauss model is assumed.
	The factor $C_{\rm Gauss}$ contains the dependence on Gauss model parameters
	and is a slowly varying function of these parameters that can be well approximated
	for practical purposes \cite{Avakian:2008dz} by its maximum
	$C_{\rm Gauss} \le C_{\rm max} = 3/(2\sqrt{2})$.
	Extractions of the (1/2)-moment of the Collins function $H_1^{\perp(1/2)}(z)=$
	$\int\di^2\vec{K}_T\frac{|\vec{K}_T|}{2zm_h}H_1^\perp(z,\vec{K}_T^{\:2})$
	from data 
	\cite{Airapetian:2004tw,Alexakhin:2005iw,Diefenthaler:2005gx,Ageev:2006da,Alekseev:2008dn}
	were reported elsewhere \cite{Vogelsang:2005cs,Efremov:2006qm,Anselmino:2007fs}.

\end{enumerate}

\newpage
\section{\boldmath Pretzelosity in the bag model}
\label{Sec-4:pretzelosity-in-bag}

In this Section we review the pretzelosity calculation \cite{Avakian:2008dz} 
in the MIT bag model, in which the nucleon consists of 3 non-interacting quarks 
confined in a spherical cavity \cite{Chodos:1974je,Celenza:1982uk,Stratmann:1993aw,Yuan:2003wk}.
The momentum space wave function is given by
\begin{equation}
    \varphi_{m}(\vec{k})=i\sqrt{4\pi}N R_0^3
    \left (\begin{array}{r} t_0(|\vec{k}|)\chi_m\\
    \vec{\sigma}\cdot\hat{k} \;t_1(|\vec{k}|)\chi_m
    \end{array} \right ) \ ,
    \label{wp}
    \end{equation}
with $N=\omega^{3/2}(2R_0^3(\omega-1)\sin^2\omega)^{-1/2}$,
$\hat{k} = \vec{k}/|\vec{k}|$. 
We fix the bag radius $R_0$ in terms of the proton mass $M_N$ as $R_0 M_N=4\omega$
with $\omega\approx 2.04$ the lowest root of the bag eigen-equation.
Finally $t_i(\kappa)=\int_0^1 u^2 du j_i(uR_0\kappa)j_i(u\omega)$
where $j_i$ are spherical Bessel functions. The bag model wave function (\ref{wp})
contains both $S$ (represented by $t_0$) and $P$ (represented by $t_1$) waves.

With the above wave functions, one obtains 
the following results \cite{Avakian:2008dz} 
\ba\label{Eq:f1-bag}
   f_1(x,k_\perp) &=& A\biggl[t_0^2+2t_0t_1\frac{k_z}{k}+t_1^2 \biggr] \\
   g_1(x,k_\perp) &=& A\biggl[t_0^2+2t_0t_1\frac{k_z}{k}+t_1^2
            (\frac{2k_z^2}{k^2}-1) \biggr]  \\
   h_1(x,k_\perp) &=& A\biggl[t_0^2+2t_0t_1\frac{k_z}{k}+t_1^2\frac{k_z^2}{k^2}\biggr]
	\\
   \label{Eq:h1Tperp-bag}
   h_{1T}^\perp(x,k_\perp) &=& A\,\biggl[-2\;\frac{M_N^2}{k^2}\;t_1^2\biggr]
\ea
with $A = 16\omega^4/[\pi^2(\omega -1)j_0^2(\omega)\,M_N^2]$ 
and the flavor dependence given by
(we assume an $SU(6)$ spin-flavor symmetric proton wave function):
\ba
	f_1^q(x,k_\perp) &=& N_q  f_1(x,k_\perp)\, , \;\; N_u = 2\, , \;\; N_d = 1\, 
	\nonumber\\ \label{Eq:wafe-function-SU(6)}
    	g_1^q(x,k_\perp) &=& P_q\,g_1(x,k_\perp)\, , \;\; P_u = \frac{4}{3}\;,\;\;
	P_d = -\frac{1}{3}\;.
\ea
The flavor dependence of $h_1^q$ and $h_{1T}^{\perp q}$ is analog to $g_1^q$.
The momenta $k_z$ and $k$ are defined as $k_z=xM_N-\omega/R_0$, and
$k=\sqrt{k_z^2+k_\perp^2}$. 

Since $h_{1T}^\perp\propto t_1^2$
it is proportional to the square of the $P$-wave~\cite{Burkardt:2007rv}
and thus sensitive to the quark orbital angular momentum in the proton.
The results (\ref{Eq:f1-bag}-\ref{Eq:wafe-function-SU(6)}) 
satisfy \cite{Avakian:2008dz} the positivity conditions in point~III,
and are consistent \cite{Avakian:2008dz} with the predictions from the 
large-$N_c$ limit discussed in point~IV of Sec.~\ref{Sec-3:What-we-know}.

From the results in Eqs.~(\ref{Eq:f1-bag}-\ref{Eq:h1Tperp-bag}) we find that
out of the 4 TMDs $f_1(x,k_\perp)$, $g_1(x,k_\perp)$, $h_1(x,k_\perp)$, 
$h_{1T}^{\perp(1)}(x,k_\perp)\equiv k_\perp^2/(2M_N^2)\,h_{1T}^{\perp}(x,k_\perp)$
only 2 are linearly independent. 
In QCD the different TMDs are, of course, all independent of each other.
But in the bag model all TMDs are expressed in terms of $t_0$ and~$t_1$ 
in Eq.~(\ref{wp}), which naturally gives rise to {\sl bag model relations} 
among different TMDs.
The most interesting (and possibly more general, see 
Sec.~\ref{Sec-6:validity-of-relation}) relation is \cite{Avakian:2008dz}
\be\label{Eq:measure-of-relativity}
	h_{1T}^{\perp(1)q}(x,k_\perp) = g_1^q(x,k_\perp) - h_1^q(x,k_\perp) \,.
\ee

%
\begin{figure}[t!]

\vspace{-0.5cm}

        \includegraphics[width=5.5cm]{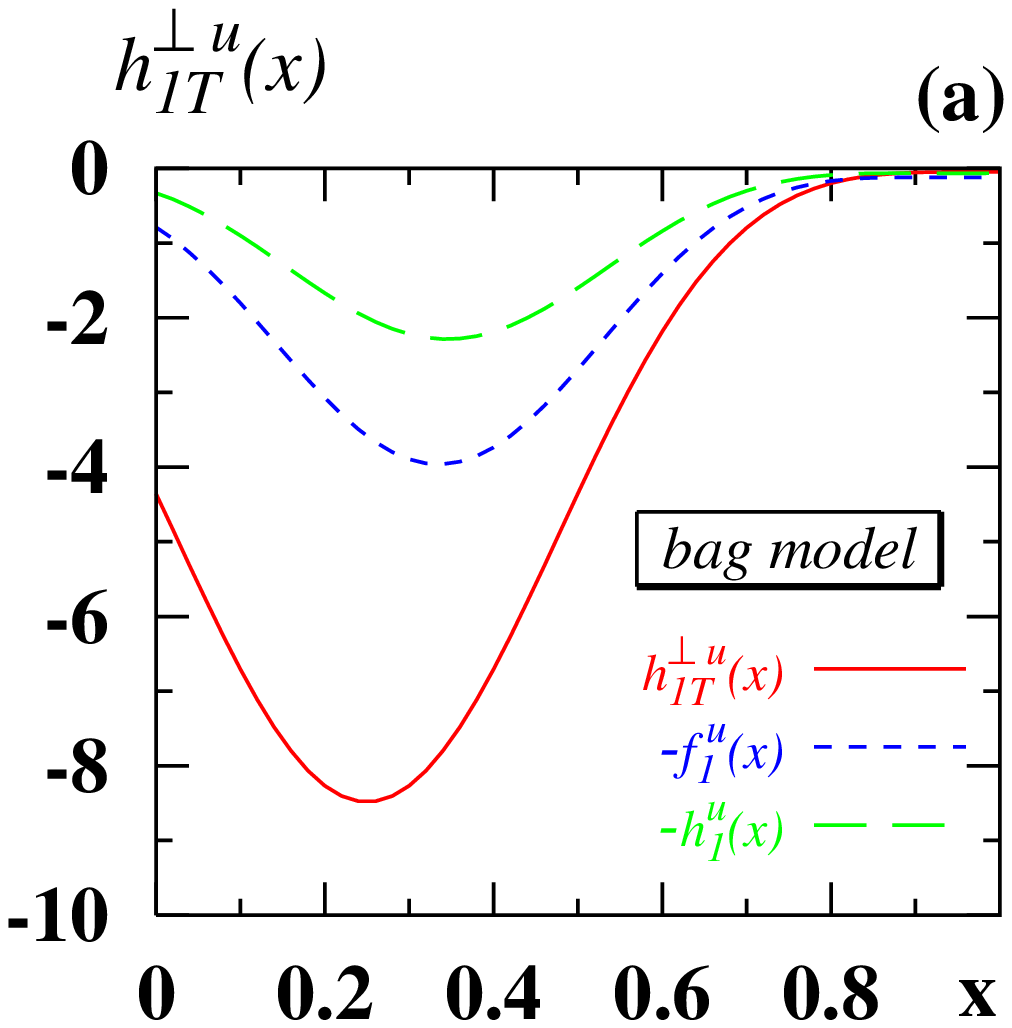}
        \includegraphics[width=5.5cm]{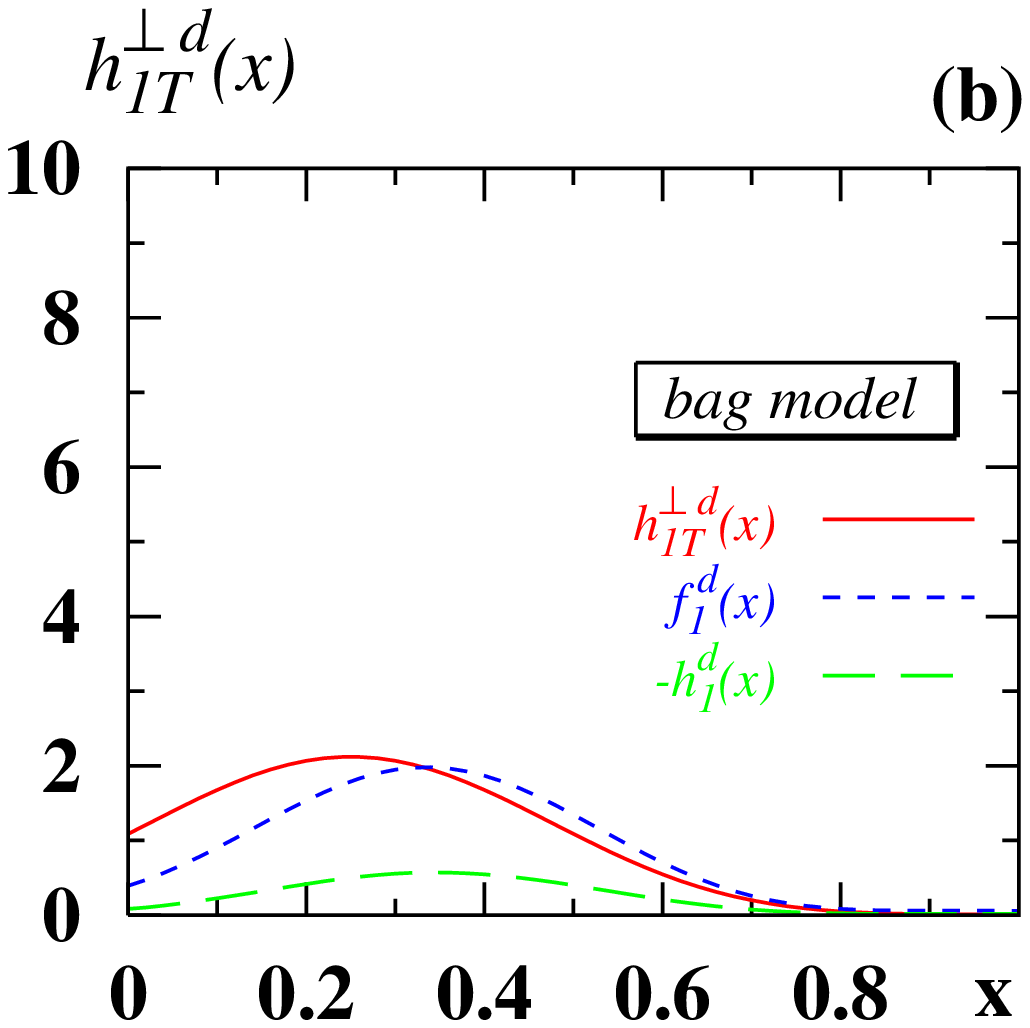}

\caption{\label{Fig02:h1Tperp-bag-x}
    The parton distribution function $h_{1T}^{\perp q}(x)$ vs.\ $x$
    from the bag model \cite{Avakian:2008dz} 
    in comparison to $f_1^q(x)$ and $h_1^q(x)$ from this model,
    The results refer to a low scale \cite{Stratmann:1993aw}. Notice that 
    $h_{1T}^{\perp q}(x)$ is not constrained by positivity.}
\end{figure}

Figs.~\ref{Fig02:h1Tperp-bag-x}a and \ref{Fig02:h1Tperp-bag-x}b show results 
for $h_{1T}^{\perp q}(x) = \int\di^2\vec{k}_\perp\,h_{1T}^{\perp q}(x,k_\perp)$
at the low bag model scale.
The pretzelosity distributions $h_{1T}^{\perp q}(x)$ have opposite 
signs compared to transversity and are larger than $h_1^q(x)$ in the bag model,
even larger than $f_1^q(x)$. However, $h_{1T}^{\perp q}(x)$  is not
constrained by positivity bounds.

	\begin{wrapfigure}[14]{HR}{5.6cm}

	\vspace{-0.2cm}

	$\phantom{X}$
   	\includegraphics[width=5.2cm]{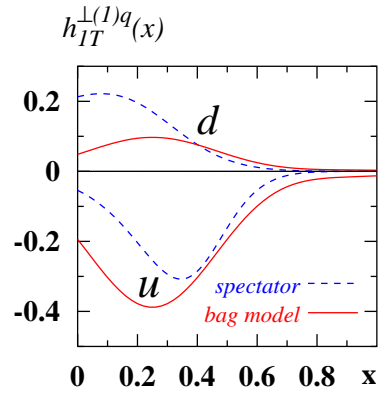}
	\caption{\label{Fig06:bag-vs-spec}
    	The $h_{1T}^{\perp (1)q}(x)$ vs.~$x$ at low scale from bag model 
	\cite{Avakian:2008dz} and spectator model \cite{Jakob:1997wg}.}
	\end{wrapfigure}

The bag \cite{Avakian:2008dz} (and spectator \cite{Jakob:1997wg}) model 
results satisfy the positivity bounds for $h_{1T}^{\perp (1)q}(x)$ 
in point~III of Sec.~\ref{Sec-3:What-we-know}.
In Fig.~\ref{Fig06:bag-vs-spec} the results for $h_{1T}^{\perp (1)q}(x)$ 
from both models are compared. We observe good qualitative agreement.

The moduli of the transverse moments $h_{1T}^{\perp (1)q}(x)$ of the
pretzelosity distribution functions are about 3 or more times 
smaller than those of the transversity distribution functions $h_1^q(x)$
with the exception of very small-$x$ where both models are strictly speaking
not applicable \cite{Avakian:2008dz}.

\section{How general is the relation in Eq.~(\ref{Eq:measure-of-relativity}) ?}
\label{Sec-6:validity-of-relation}

Eq.~(\ref{Eq:measure-of-relativity}) is remarkable from the point of view 
of the observation that \cite{Jaffe:1991ra} 
'the difference of $g_1^q(x)$ and $h_1^q(x)$ 
is a measure for relativistic effects in nucleon'. 
This difference is just the transverse moment of pretzelosity!

It is clear that this relation cannot be strictly valid in QCD,
where all TMDs are independent. However, could it nevertheless
allow to gain a reasonable estimate for $h_{1T}^{\perp(1)}(x)$ 
in terms of transversity and helicity?
Until clarified by experiment, we can address this question only in models.

Interestingly, the relation (\ref{Eq:measure-of-relativity}) does not hold 
only in the bag model \cite{Avakian:2008dz}, but is found  \cite{Avakian:2008dz} 
to be satisfied also in the spectator model \cite{Jakob:1997wg}. In fact, 
it was conjectured \cite{Avakian:2008dz} that (\ref{Eq:measure-of-relativity}) 
could hold in a large class of {\sl relativistic quark models}, 
which was subsequently confirmed in the constituent quark model 
\cite{Pasquini:2008ax} and the relativistic model of the proton \cite{pretzel-new}. 
But  (\ref{Eq:measure-of-relativity}) is not satisfied in a different than 
\cite{Jakob:1997wg} version of the spectator model \cite{Bacchetta:2008af}.

The limitations of (\ref{Eq:measure-of-relativity}) are nicely illustrated
in the 'quark target model' where in addition to quarks
there are also gluons, and
(\ref{Eq:measure-of-relativity}) is not satisfied \cite{Meissner:2007rx}.
Thus, the explicit inclusion of gluon degrees of freedom spoils this relation.
Of course, as stressed above, we do not expect (\ref{Eq:measure-of-relativity})
to be valid in QCD.

It would be interesting to know the necessary and sufficient
conditions for the relation (\ref{Eq:measure-of-relativity})
to hold in a model.

%
\begin{figure}[b!]
\begin{tabular}{cc}

    \hspace{0.5cm}\includegraphics[width=5cm]{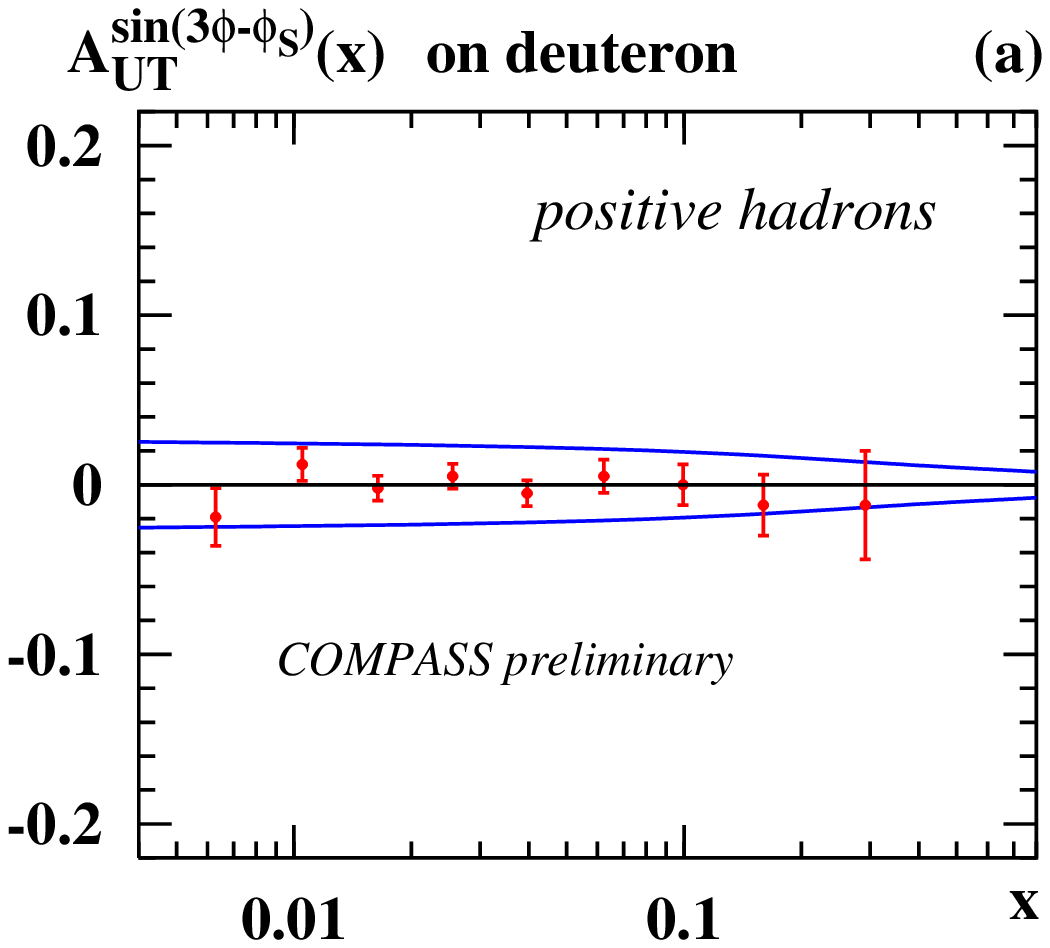} &
    \hspace{-0.5cm}\includegraphics[width=5cm]{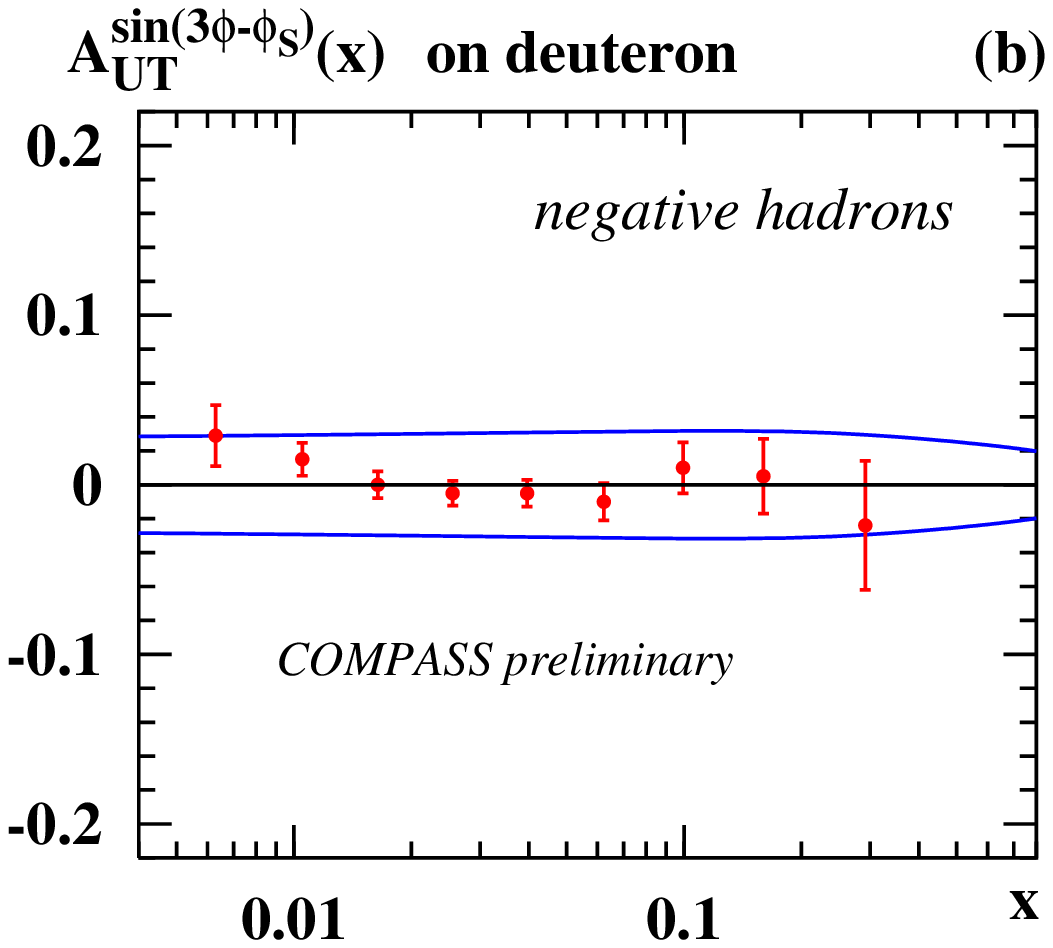}
\end{tabular}
\caption{\label{Fig03:AUT-deut-COMPASS-x}
    The transverse target SSA $A_{UT,\,\pi}^{\sin(3\phi-\phi_S)}$
    for deuteron estimated on the basis of the positivity bound vs.\
    preliminary COMPASS data \cite{Kotzinian:2007uv}.}
\end{figure}
%

\section{\boldmath Preliminary COMPASS data \& prospects at JLab}

At COMPASS the $\sin(3\phi-\phi_S)$ and other SSAs
were measured on a deuteron target \cite{Kotzinian:2007uv}.
By saturating the positivity bound for $h_{1T}^{\perp(1)}(x)$
(point~III in Sec.~\ref{Sec-3:What-we-know}) we estimated \cite{Avakian:2008dz} 
the maximum effect for $A_{UT}^{\sin(3\phi-\phi_S)}$. For that
information on Collins effect \cite{Efremov:2006qm,Anselmino:2007fs} 
and parameterizations \cite{Gluck:1998xa,Kretzer:2000yf} were used.
The results are shown in Fig.~\ref{Fig03:AUT-deut-COMPASS-x}
and compared to the preliminary data \cite{Kotzinian:2007uv}.

%
	\begin{wrapfigure}[16]{HR}{5.6cm}

	\vspace{-0.8cm}

    \includegraphics[width=5.6cm]{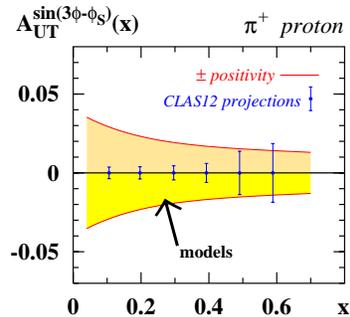}
\caption{\label{Fig9:pretzel-at-CLAS}
    SSA $A_{UT}^{\sin(3\phi-\phi_S)}$ in $\pi^+$ production
    from proton in the kinematics of CLAS 12 as function of $x$
    (error projections from \cite{Avakian-LOI-CLAS12}).
    The shaded areas indicate the range allowed by positivity.
    The bag \cite{Avakian:2008dz} and spectator \cite{Jakob:1997wg}
    models predict a negative SSA.}
\end{wrapfigure}

At small $x$ the preliminary data favor that 
pretzelosity does not reach the bound. 
Whether due to the expected suppression at small $x$ or opposite signs 
of $u$ and $d$-flavors, see Sec.~\ref{Sec-3:What-we-know}, cannot 
be concluded.

The important observation is that preliminary COMPASS data 
\cite{Kotzinian:2007uv} do not exclude a sizeable effect 
in the region $x>0.1$, see Fig.~\ref{Fig03:AUT-deut-COMPASS-x},
where JLab can measure with precision.
This is demonstrated in Fig.~\ref{Fig9:pretzel-at-CLAS} showing the $\pi^+$
SSA from a proton target in the kinematics of CLAS with 12 GeV beam upgrade
(with error projections for 2000 hours run time \cite{Avakian-LOI-CLAS12}). 

It could be even more promising to look at SSAs due to Collins 
effect, like $A_{UT}^{\sin(3\phi-\phi_S)}$, in kaon production.
The statistics for kaon production is lower than for pion production, 
but the SSA might be larger as it is suggested by a model
\cite{Bacchetta:2007wc} of the Collins function.
With a RICH detector at CLAS the kaon SSAs could be measured in the
valence-$x$ region \cite{CLAS-12-RICH}.

\section{Conclusions}

We reviewed the bag model calculation of pretzelosity \cite{Avakian:2008dz}. 
An interesting observation is the relation 
(\ref{Eq:measure-of-relativity}) which connects helicity and 
transversity distributions to $h_{1T}^{\perp(1)}(x)$, 
and is valid in many 
\cite{Jakob:1997wg,Avakian:2008dz,Pasquini:2008ax,pretzel-new,Meissner:2007rx}
(though not all \cite{Meissner:2007rx,Bacchetta:2008af}) models,
but not in QCD where all TMDs are independent \cite{Goeke:2005hb}.
Nevertheless (\ref{Eq:measure-of-relativity}) could turn out a useful 
approximation. In view of the numerous novel TMDs, well-motivated
approximations are welcome \cite{Avakian:2007mv}.

In the bag model  $h_{1T}^{\perp u}$ is proportional to {\sl minus} 
the square of the $p$-wave component of the nucleon wave function,
and therefore manifestly negative ($h_{1T}^{\perp d}$ has
opposite sign dictated by SU(6) symmetry, and large $N_c$ \cite{Pobylitsa:2003ty}). 
As a consequence of (\ref{Eq:measure-of-relativity})
we have $|h_1^q(x)|>|g_1^q(x)|$. This is found in models 
\cite{Schweitzer:2001sr,Efremov:2004tz}.

Forthcoming analyzes and experiments at COMPASS, HERMES and JLab
\cite{Avakian-LOI-CLAS12,Avakian-clas6,Avakian-clas12}
will provide valuable information on the pretzelosity distribution
function, and deepen our understanding of the nucleon structure.

\vspace{0.5cm}

\noindent{\bf Acknowledgements.}
We thank the organizers of the {\sl Transversity~2008}
for the opportunity to present our work supported by:
BMBF,
German-Russian collaboration (DFG-RFFI, 436 RUS 113/881/0),
the EIIIHP project 
RII3-CT-2004-506078, 
the Grants RFBR 06-02-16215 and 07-02-91557,
RF MSE RNP.2.2.2.2.6546 (MIREA), 
the Heisenberg-Landau Program of JINR,
the contract DE-AC05-06OR23177, 
RIKEN, Brookhaven National Laboratory, and 
the U.S.\ Department of Energy DE-AC02-98CH10886.



\begin{thebibliography}{99}


\bibitem{Collins:1981uk}
  J.~C.~Collins and D.~E. Soper,
  Nucl.\ Phys.\ B {\bf 193}, 381 (1981)
  [Erratum-ibid.\ B {\bf 213}, 545 (1983)]. 

\bibitem{Collins:1984kg}
  J.~C.~Collins, D.~E.~Soper and G.~Sterman,
  Nucl.\ Phys.\ B {\bf 250}, 199 (1985).

\bibitem{Ji:2004wu}
  X.~D.~Ji, J.~P.~Ma and F.~Yuan,
  Phys.\ Rev.\ D {\bf 71}, 034005 (2005).\\
  X.~D.~M.~Ji, J.~P.~M.~Ma and F.~Yuan,
  Phys.\ Lett.\ B {\bf 597}, 299 (2004).

\bibitem{Collins:2004nx}
  J.~C.~Collins and A.~Metz,
  Phys.\ Rev.\ Lett.\  {\bf 93}, 252001 (2004).

\bibitem{Boer:1997nt}
  D.~Boer and P.~J.~Mulders,
  Phys.\ Rev.\ D {\bf 57}, 5780 (1998).
\bibitem{Mulders:1995dh}
  P.~J.~Mulders and R.~D.~Tangerman,
  Nucl.\ Phys.\ B {\bf 461} (1996) 197
  [Erratum-ibid.\ B {\bf 484} (1997) 538].

\bibitem{Brodsky:2002cx}
  S.~J.~Brodsky, D.~S.~Hwang and I.~Schmidt,
  Phys.\ Lett.\ B {\bf 530}, 99 (2002);
  Nucl.\ Phys.\ B {\bf 642}, 344 (2002).

\bibitem{Collins:2002kn}
  J.~C.~Collins,
  Phys.\ Lett.\ B {\bf 536}, 43 (2002) [arXiv:hep-ph/0204004].

\bibitem{Belitsky:2002sm}
  A.~V.~Belitsky, X.~Ji and F.~Yuan,
  Nucl.\ Phys.\ B {\bf 656}, 165 (2003).\\
  X.~D.~Ji and F.~Yuan,
  Phys.\ Lett.\ B {\bf 543}, 66 (2002).
  \\
  D.~Boer, P.~J.~Mulders and F.~Pijlman,
  Nucl.\ Phys.\ B {\bf 667}, 201 (2003)

\bibitem{Cherednikov:2007tw}
  I.~O.~Cherednikov and N.~G.~Stefanis,
  Phys.\ Rev.\  D {\bf 77}, 094001 (2008);
  I.~O.~Cherednikov and N.~G.~Stefanis,
  Nucl.\ Phys.\  B {\bf 802}, 146 (2008).

\bibitem{Burkardt:2007rv}
  M.~Burkardt,
  arXiv:0709.2966 [hep-ph].

\bibitem{Miller:2007ae}
  G.~A.~Miller,
  Phys.\ Rev.\  C {\bf 76}, 065209 (2007).

\bibitem{Avakian:2007xa}
  H.~Avakian, S.~J.~Brodsky, A.~Deur and F.~Yuan,
  Phys.\ Rev.\ Lett.\  {\bf 99}, 082001 (2007).

\bibitem{Brodsky:2006hj}
  S.~J.~Brodsky and F.~Yuan,
  Phys.\ Rev.\  D {\bf 74} (2006) 094018.

\bibitem{Bacchetta:1999kz}
  A.~Bacchetta, M.~Boglione, A.~Henneman and P.~J.~Mulders,
  Phys.\ Rev.\ Lett.\  {\bf 85}, 712 (2000).


\bibitem{Soffer:1994ww}
  J.~Soffer,
  Phys.\ Rev.\ Lett.\  {\bf 74} (1995) 1292.

\bibitem{Pobylitsa:2003ty}
  P.~V.~Pobylitsa,
  arXiv:hep-ph/0301236.

\bibitem{Avakian:2008dz}
  H.~Avakian, A.~V.~Efremov, P.~Schweitzer and F.~Yuan,
  arXiv:0805.3355. 

\bibitem{Jakob:1997wg}
  R.~Jakob, P.~J.~Mulders and J.~Rodrigues,
  Nucl.\ Phys.\  A {\bf 626}, 937 (1997).

\bibitem{Pasquini:2008ax}
  B.~Pasquini, S.~Cazzaniga and S.~Boffi,
  arXiv:0806.2298 [hep-ph].\\
  B.~Pasquini,
  arXiv:0807.2825 [hep-ph].

\bibitem{pretzel-new}
  A.~V.~Efremov, P.~Schweitzer, O.~V.~Teryaev and P.~Zavada,
  in preparation
  (P.~Zavada, talk at {\sl SPIN Praha 2008}, Prague, 20-26 July 2008).

\bibitem{Meissner:2007rx}
  S.~Meissner, A.~Metz and K.~Goeke,
  Phys.\ Rev.\  D {\bf 76}, 034002 (2007).

\bibitem{Bacchetta:2008af}
  A.~Bacchetta, F.~Conti and M.~Radici,
  arXiv:0807.0323 [hep-ph].

\bibitem{Collins:1992kk}
  J.~C.~Collins,
  Nucl.\ Phys.\ B {\bf 396}, 161 (1993). 

\bibitem{Efremov:1992pe}
  A.~V.~Efremov, L.~Mankiewicz and N.~A.~Tornqvist,
  Phys.\ Lett.\ B {\bf 284} (1992) 394.
  J.~C.~Collins, S.~F.~Heppelmann and G.~A.~Ladinsky,
  Nucl.\ Phys.\ B {\bf 420} (1994) 565.


\bibitem{Airapetian:2004tw}
  A.~Airapetian {\it et al.}  [HERMES Coll.],
  Phys.\ Rev.\ Lett.\  {\bf 94}, 012002 (2005).

\bibitem{Alexakhin:2005iw}
  V.~Y.~Alexakhin {\it et al.}  [COMPASS Coll.],
  Phys.\ Rev.\ Lett.\  {\bf 94}, 202002 (2005).

\bibitem{Diefenthaler:2005gx}
  M.~Diefenthaler [HERMES Coll.],
  AIP Conf.\ Proc.\  {\bf 792} (2005) 933,
  arXiv:0706.2242 
  and
  arXiv:hep-ex/0612010.

\bibitem{Ageev:2006da}
  E.~S.~Ageev {\it et al.}  [COMPASS Coll.],
  Nucl.\ Phys.\  B {\bf 765} (2007) 31

\bibitem{Alekseev:2008dn}
  M.~Alekseev {\it et al.}  [COMPASS Coll.],
  arXiv:0802.2160 [hep-ex].

\bibitem{Vogelsang:2005cs}
  W.~Vogelsang and F.~Yuan,
  Phys.\ Rev.\ D {\bf 72} (2005) 054028.

\bibitem{Efremov:2006qm}
  A.~V.~Efremov, K.~Goeke and P.~Schweitzer,
  Phys.\ Rev.\  D {\bf 73}, 094025 (2006).

\bibitem{Anselmino:2007fs}
  M.~Anselmino, M.~Boglione, U.~D'Alesio, A.~Kotzinian, F.~Murgia, A.~Prokudin
  and C.~Turk,
  Phys.\ Rev.\  D {\bf 75}, 054032 (2007).

\bibitem{Chodos:1974je}
  A.~Chodos, R.~L.~Jaffe, K.~Johnson, C.~B.~Thorn and V.~F.~Weisskopf,
  Phys.\ Rev.\  D {\bf 9}, 3471 (1974).
  R.~L.~Jaffe,
  Phys.\ Rev.\  D {\bf 11}, 1953 (1975).

\bibitem{Celenza:1982uk}
  L.~S.~Celenza and C.~M.~Shakin,
  Phys.\ Rev.\  C {\bf 27}, 1561 (1983)
  [Erratum-ibid.\  C {\bf 39}, 2477 (1989)].

\bibitem{Stratmann:1993aw}
  M.~Stratmann,
  Z.\ Phys.\  C {\bf 60}, 763 (1993).

\bibitem{Yuan:2003wk}
  F.~Yuan,
  Phys.\ Lett.\  B {\bf 575}, 45 (2003).

\bibitem{Jaffe:1991ra}
  R.~L.~Jaffe and X.~D.~Ji,
  Nucl.\ Phys.\  B {\bf 375} (1992) 527, and
  Phys.\ Rev.\ Lett.\  {\bf 67}, 552 (1991).

\bibitem{Kotzinian:2007uv}
  A.~Kotzinian  [on behalf of the COMPASS Coll.],
  arXiv:0705.2402 [hep-ex].

\bibitem{Gluck:1998xa}
  M.~Gl\"uck, E.~Reya and A.~Vogt,
  Eur.\ Phys.\ J.\ C {\bf 5}, 461 (1998).
  M.~Gl\"uck, E.~Reya, M.~Stratmann and W.~Vogelsang,
  Phys.\ Rev.\ D {\bf 63}, 094005 (2001).

\bibitem{Kretzer:2000yf}
  S.~Kretzer,
  Phys.\ Rev.\ D {\bf 62} (2000) 054001.

\bibitem{Avakian-LOI-CLAS12}
  H.~Avakian,  {\it et al.}, JLab LOI 12-06-108 (2008).

\bibitem{Bacchetta:2007wc}
  A.~Bacchetta, L.~P.~Gamberg, G.~R.~Goldstein and A.~Mukherjee,
  Phys.\ Lett.\  B {\bf 659}, 234 (2008)
  [arXiv:0707.3372 [hep-ph]].

\bibitem{CLAS-12-RICH}
  {\sl CLAS 12 Rich Detector Workshop}, 
  28-29 January 2008, Jefferson Lab, Newport News, VA, U.S.A.
  {\tt http://conferences.jlab.org/CLAS12}

\bibitem{Goeke:2005hb}
  K.~Goeke, A.~Metz and M.~Schlegel,
  Phys.\ Lett.\ B {\bf 618}, 90 (2005).

\bibitem{Avakian:2007mv}
  H.~Avakian, A.~V.~Efremov, K.~Goeke, A.~Metz, P.~Schweitzer and T.~Teckentrup,
  Phys.\ Rev.\  D {\bf 77}, 014023 (2008).

\bibitem{Schweitzer:2001sr}
  P.~Schweitzer {\it et al.},
  Phys.\ Rev.\ D {\bf 64} (2001) 034013.\\
  K.~Goeke {\it et al.},
  Acta Phys.\ Polon.\ B {\bf 32}, 1201 (2001).\\
  M.~Wakamatsu and T.~Kubota,
  Phys.\ Rev.\ D {\bf 60}, 034020 (1999).

\bibitem{Efremov:2004tz}
  A.~V.~Efremov, O.~V.~Teryaev and P.~Zavada,
  Phys.\ Rev.\  D {\bf 70} (2004) 054018.\\
  P.~Zavada,
  Eur.\ Phys.\ J.\  C {\bf 52}, 121 (2007),
  Phys.\ Rev.\  D {\bf 55}, 4290 (1997),
  Phys.\ Rev.\  D {\bf 67}, 014019 (2003).

\bibitem{Avakian-clas6}
  H.~Avakian  {\it et al.}, JLab E05-113,
  ``Semi-Inclusive Pion Production with a Longitudinally Polarized Target at 6 GeV''.

\bibitem{Avakian-clas12}
  H.~Avakian  {\it et al.}, JLab PR12-07-107,
  ``Studies of Spin-Orbit Correlations with Longitudinally Polarized Target''.



\end{thebibliography}
\end{document}